\documentclass[multphys,vecphys]{svmult}

\usepackage{makeidx}     
\usepackage{graphicx}    
\usepackage{multicol}    

\makeindex             

\begin{document}

\title*{Optical and near-IR observations of SN~1998bw}
 \author{Ferdinando Patat}
\institute{European Southern Observatory
\texttt{fpatat@eso.org}}
%
%
\maketitle

\abstract
SN~1998bw, especially after the discovery of GRB~030329/SN~2003dh,
seems to be the equivalent of the Rosetta stone for the SN/GRB
connection. In this paper I review optical and near IR observations
that have been carried out for this uncanny object, which has probably
confirmed suspicions and ideas originally formulated in the early
seventies of last century.

\vspace{5mm}
\begin{quote}
{\it \footnotesize Thus, the observation of gamma-ray line emission
from a young supernova seems very promising in the near future. The
observation, or even a null observation at a low threshold, will give
significance in the fields of nuclear astrophysics and supernova
theory. The scientific importance of a positive measurement would be
analogous with and comparable to the importance of successful
detection of neutrinos from the Sun.}

\vspace{5mm}
\hspace{60mm}Clayton, Colgate \& Fishman \cite{clayton}.
\end{quote}

\vspace{10mm}

This story probably begins in 1969, with what I like to call a
prophecy, and it is right with it that I wish to start this review on
the optical and near-IR observations of SN~1998bw\footnote{This talk was 
given in Valencia on April 25, 2003. For some cabalistic reason, this 
coincided exactly with the fifth anniversary of GRB~980425/SN1998bw.}.

As J. Sollerman said in one of his papers on this striking object, SN~1998bw was 
born famous. And it was doomed to become even more famous as time went by, so famous
that it was sometimes named {\it the supernova of the century}.
And this is indeed interesting, since it was born in the same century of SN~1987A,
one of the most studied and referenced objects in the sky.

Just from the optical and near-IR observations point of view, this is witnessed
by the large number of papers which have been published in the first four years.
Starting with the {\it Nature} papers by Galama et al. \cite{galama} and 
Iwamoto et al. \cite{iwamoto}, a number of authors have presented the results of
their observational campaigns: McKenzie \& Schaefer \cite{mckenzie}, Galama et al.
\cite{galama99}, Stathakis et al. \cite{stathakis}, Fynbo et al. \cite{fynbo},
Sollerman et al. \cite{sollerman00}, Patat et al. \cite{patat01} and Sollerman 
et al. \cite{sollerman02}.

The reader is referred to these papers for a detailed account on the observations, while
here I will try to give only a general view of the SN~1998bw phenomenon.

\begin{figure}
\centering
\includegraphics[height=10cm]{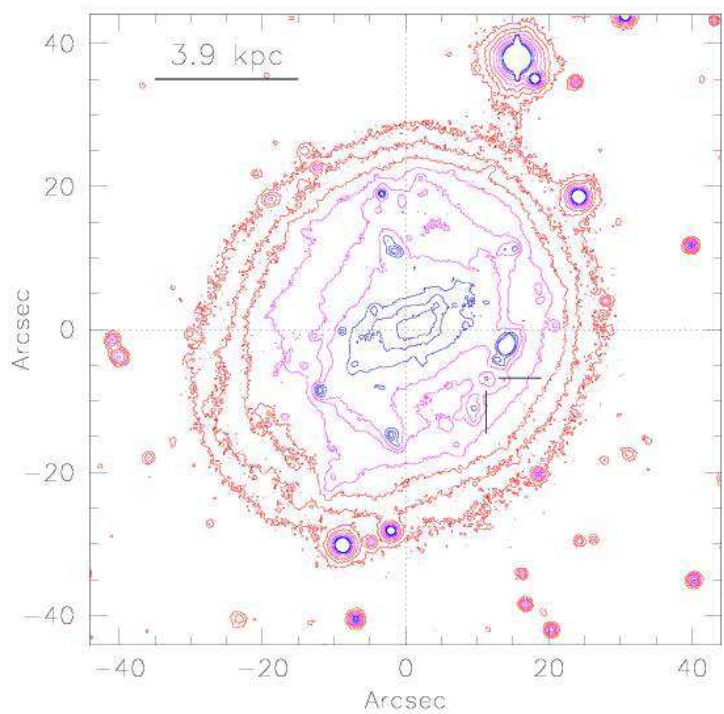}
\caption{Contour plot of ESO-184-G82. The original image stack was obtained
at VLT+FORS1 at about 900 days after the explosion. Spatial scale was computed
for a distance of 40 Mpc. The SN position is marked.}
\label{fig:contour}
\end{figure}

SN~1998bw was discovered by Galama et al. \cite{galama98a} in the
BeppoSAX Wide Field Camera error box of GRB~980425 (Soffita et
al. \cite{soffita}, Pian et al. \cite{pian}) close to a spiral arm of
the barred galaxy ESO~184--G82 (see Fig.~\ref{fig:contour}), by
comparing two frames taken at the ESO New Technology Telescope
on Apr 28.4 and May 1.3 UT.  Spectroscopic and photometric
observations, both in the optical and in the near IR, started at
ESO--La Silla immediately after the discovery, and showed that this
object was profoundly different from all then known SNe (Lidman et
al. \cite{lidman}).

\begin{figure}
\centering
\includegraphics[height=10cm]{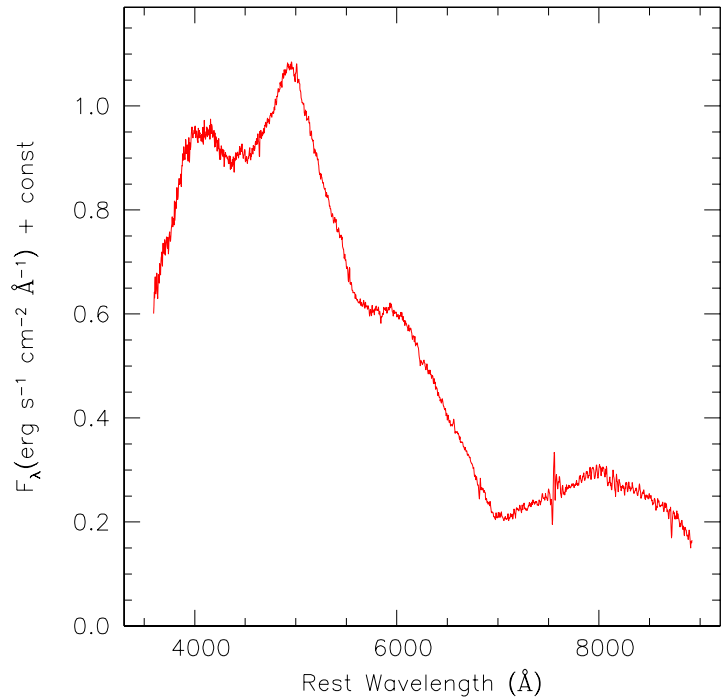}
\caption{Spectrum of SN~1998bw taken at ESO-La Silla on May 5, 1998.}
\label{fig:firstspec}
\end{figure}

Its peculiar spectroscopic appearance, its unusually high radio
luminosity at early phases (Kulkarni et al. \cite{kulkarni}), its
optical luminosity ($M_V\sim -$19.2 + 5 log $h_{65}$) and, in
particular, the probable association with GRB~980425 through
positional and temporal coincidence (Galama et al. \cite{galama98a},
Pian et al. \cite{pian}) placed SN~1998bw at the center of discussion
concerning the nature of Gamma Ray Bursts. The object was tentatively
classified as a peculiar Ic (Patat \& Piemonte \cite{patat98a},
Filippenko \cite{filippenko98}), I would say {\it by definition} more
than anything else, due to the complete absence of H lines, the
weakness of the Si~II 6355~\AA\/ line and no clear He~I detection in
the optical spectra (see Fig.~\ref{fig:firstspec}).  Main spectral
features were identified as O~I, Ca~II, Si~II and Fe~II
(Iwamoto et al. \cite{iwamoto}). The estimated expansion velocities
were exceptionally high ($\sim$30,000 km s$^{-1}$) and this caused a
severe line blending. The evolution during the first months was
unusually slow compared to known Ic, with the nebular spectra still
retaining many of the features present during the photospheric phase
(Stathakis et al. \cite{stathakis}, Patat et al. \cite{patat01}).

The late onset of the fully 
nebular phase has been interpreted as an indication for a large ejected mass 
(Stathakis et al. \cite{stathakis}) as it was predicted by the early light curve models. During 
the intermediate phase, the emission lines were definitely broader than in known 
Type Ib/c SNe and the simultaneous presence of iron--peak
and $\alpha$-elements indicated unusual relative abundances or physical
conditions in the SN ejecta (Patat et al. \cite{patat01}).
The late spectroscopy presented by Sollerman et al. \cite{sollerman00} showed that
the tentative morphological classification of SN~1998bw as a Type Ic 
event was indeed appropriate. The main features have been identified as
[O~I], Ca~II, Mg~I and Na~I~D, the latter possibly contaminated by He~I~5876~\AA.

As far as the Gamma-ray burst is concerned, GRB~980425 was pretty
weak, since the implied energy for a 40 Mpc distance was 8.1$\pm$1.0
$\times$10$^{47}$ erg (Pian et al. \cite{pian}), which is definitely
smaller than the usual 10$^{53}$ erg value typical for the so-called
cosmological GRBs. This has led the community to believe that GRB~980425
is a member of an unusual class of GRBs/SNe (see for example Fynbo et al. 
\cite{fynbo}).

The galaxy which hosted GRB~980425/SN~1998bw, ESO~184--G82, is an SBc
galaxy with a recession velocity $v_r$=2532 km s$^{-1}$ (Patat et
al. \cite{patat01}). Its luminosity is $L\sim$0.5$-$1.2 $L_{LMC}$,
it is currently undergoing strong star formation, it shows a bar
and rather clear indications of morphological disturbances 
(Fynbo et al.\cite{fynbo}). This is clearly visible in the late VLT
images, which show a possibly double nucleus, isophotal twisting and 
asymmetry (see Fig.~\ref{fig:contour}). All this suggests that the
observed star formation is related to galaxy interaction/merging.

The HST images have shown that SN~1998bw exploded in a star-forming
region (Fynbo et al.\cite{fynbo}), containing several bright and young
stars within a projected distance of 100 pc. This is consistent with
the progenitor of SN~1998bw being a young and massive star

Due to its relatively high apparent brightness, Kay et al. \cite{kay}
and Patat et al. \cite{patat01} have attempted to perform some
polarimetric measurements at different epochs. 
After correcting for the interstellar polarization in the direction
of the host galaxy, the estimated optical linear polarization was 0.6\%
(day $-$7), 0.4\% (day +10) and 0.5\% (day +42).
The fluctuations in these values seem to suggest that the observed
polarization is intrinsic to the SN, although a dusty medium
in the parent galaxy cannot be ruled out.

\begin{figure}
\centering
\includegraphics[height=12cm,angle=-90]{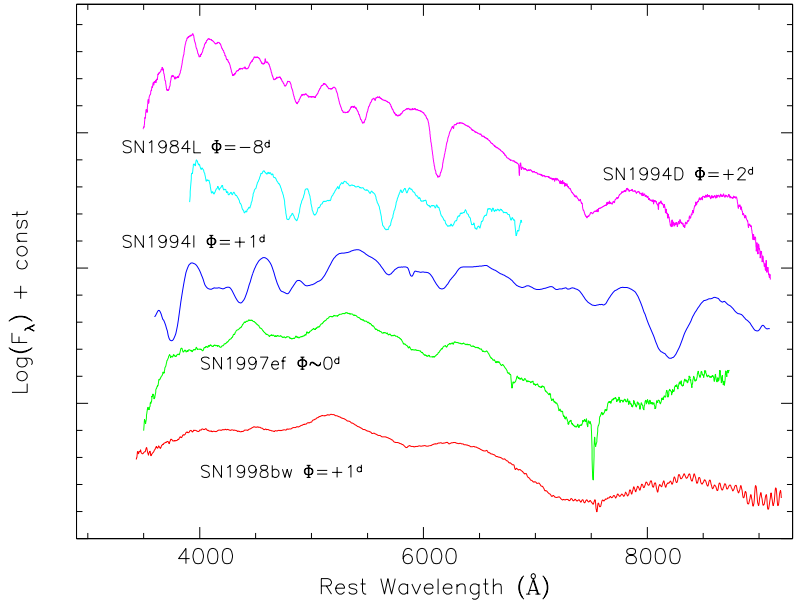}
\caption{Comparison between SN~1998bw and other SNe at maximum light.}
\label{fig:compmax}
\end{figure}

The small degree of polarization at optical wavelengths can be explained
in terms of a moderate departure from sphericity (axial ratio less
than 2:1; H\"oflich et al. \cite{hoflich}), either in the photosphere
or in the outer scattering envelope when the line of sight is not coincident 
with an axis of symmetry.


Interestingly, no polarization, either circular or linear, was detected
in the radio, and this has been interpreted as the signature of a
spherically symmetric blast wave (Kulkarni et al. \cite{kulkarni}).
This apparent discrepancy might suggest that the radio and the optical
radiation were generated in regions of different geometry.

\section{Photometric and spectroscopic evolution}

The early light curve of SN~1998bw has shown that the object was unusually
bright when compared to known SNe of type Ib/c (Galama et al. \cite{galama})
and, in this respect, it was much more similar to a type Ia. The broad-band
photometric observations by McKenzie \& Schaefer \cite{mckenzie} taken during
the intermediate phases revealed that the object settled on an exponential
decay similar to that observed in other type Ic SNe. McKenzie \& Schaefer
first suggested that even in this case the light curve was powered by the
radioactive decay of $^{56}$Co with some leakage of $\gamma$-rays.
Photometry covering later phases was then presented by Sollerman et al.
\cite{sollerman00}, Patat et al. \cite{patat01} and Sollerman et al.
\cite{sollerman02}, the latter extending to $\sim$1000 days after the
explosion by means of HST observations.

The late light curve continues to fall significantly steeper that the
decay rate of $^{56}$Co up to more than 500 days. There is no sign of
the so-called positron phase, in which the fully deposited kinetic
energy from the positrons would dominate the light curve.

Another interesting feature is the light curve flattening observed at
about 800 days past explosion. For a detailed discussion on the
possible explanations, we refer the reader to the original paper by
Sollerman et al. \cite{sollerman02} and here we just mention them:
onset of more long-lived isotopes radioactive decay (e.g. $^{57}Co$),
freeze-out, interaction with CSM, black-hole powering and faint
light echoes.

Using a simple radioactive model, Sollerman et al. \cite{sollerman02}
could fit the data with $\sim$0.3 M$_\odot$ of ejected $^{56}$Ni,
which can be regarded as a lower limit to the amount of ejected
nickel in SN~1998bw. In this respect we note that the
early light curve modeling (see Iwamoto et al. \cite{iwamoto})
required $\sim$0.7 M$_\odot$ in order to reproduce the observed
high peak luminosity.

\begin{figure}
\centering
\includegraphics[height=12cm,angle=-90]{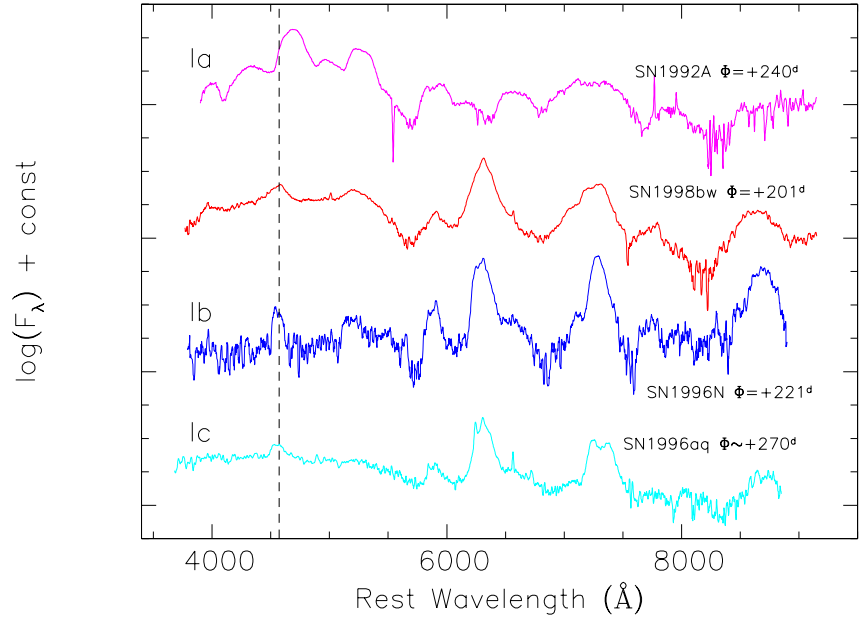}
\caption{Comparison between SNe 1992A (Ia), 1998bw, 1991N (Ib) and 1996aq (Ic)
at late phases. The vertical dashed line is placed at the rest-frame wavelength of
MgI]$\lambda$4571.}
\label{fig:complate}
\end{figure}

\begin{figure}
\centering
\includegraphics[height=12cm,angle=-90]{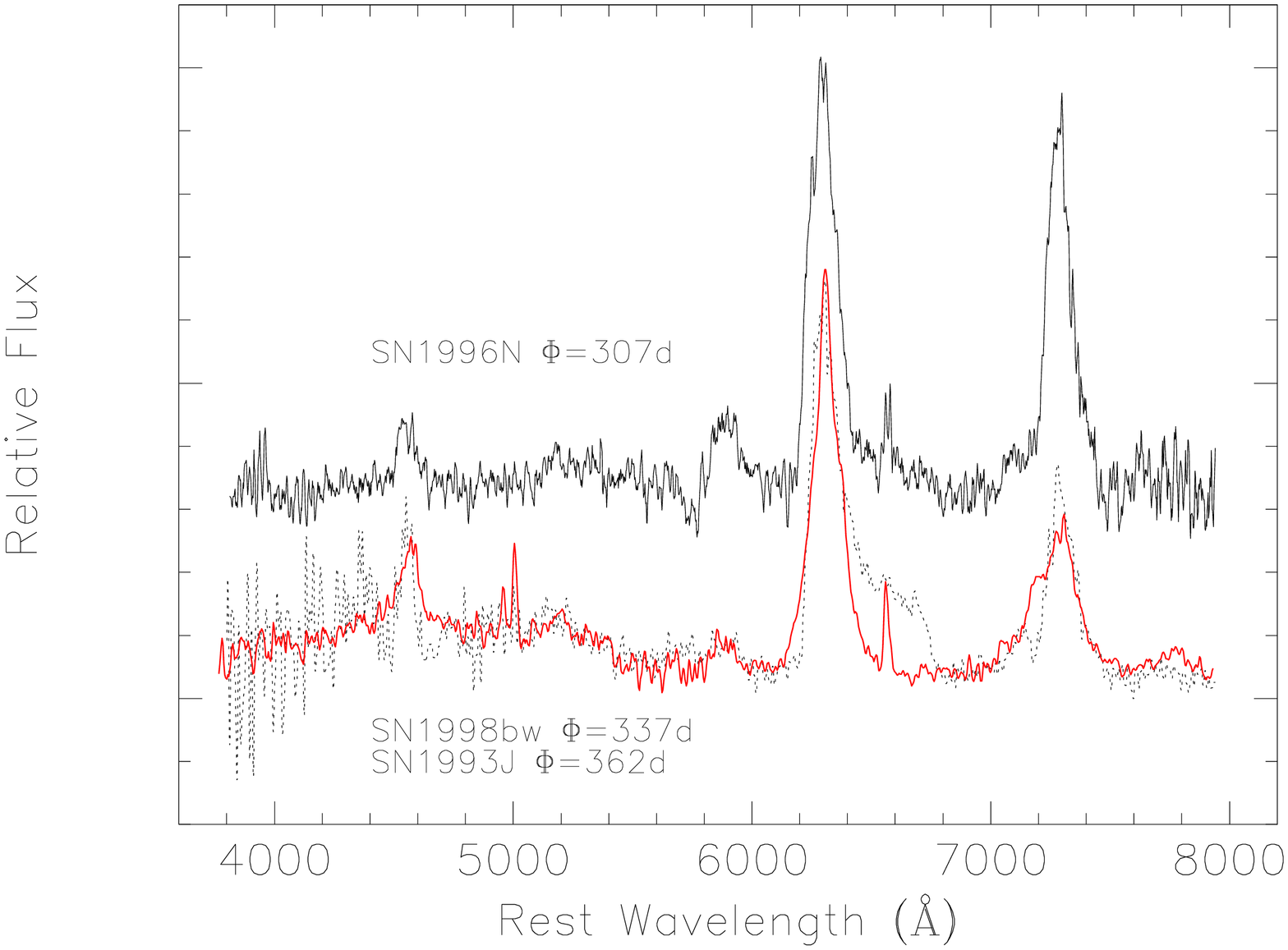}
\caption{Comparison between Type IIb SN~1993J (dotted line), Type Ib 1996N and
SN~1998bw at about 1 year after maximum light. Spectra have been normalized to the
[OI]$\lambda\lambda$6300,6364 peak and arbitrarily shifted for presentation.}
\label{fig:final}
\end{figure}

Extensive spectroscopic data sets were presented by Stathakis et al.
\cite{stathakis} and Patat et al. \cite{patat01}.  The general
appearance of the spectrum at maximum light is quite unique among SNe,
even though it is somewhat reminiscent of SN~1997ef (see
Fig.~\ref{fig:compmax}), which has been modeled as a massive SN Ic
(Mazzali et al. \cite{mazzali}).  At these early epochs, when the
velocity is high, line blending is particularly severe; the modeling
presented by Iwamoto et al.  \cite{iwamoto} suggests that the main
features are due to lines of Si~II, O~I, Ca~II and Fe~II.
The velocity, deduced from the Si~II $\lambda$6355 line is about
30,000 km s$^{-1}$ at day $-$7 and decreases to about 18,000 km s$^{-1}$
at day $+$22. These value are exceptionally high, for any SN.

Starting at about one month after maximum light, the SN enters its
nebular phase. The transition from an absorption to an emission
spectrum is slow and subtle. While the evolution of SN~1998bw in the
range 5500-9000 \AA\/ is similar to that of SN Ic events, the
expansion velocities are larger, and the region between 4000 and 5500
\AA\/ is dominated (at least until about day +200) by a wide bump to
which Fe~II transitions probably contribute significantly (see
Fig.~\ref{fig:complate}).  In general, the spectral appearance
supports the idea that this object is related to SNe Ib/c. It might be
regarded as an extreme case among these objects, having large kinetic
energy, ejecta mass and ejected mass of synthesized $^{56}$Ni, while
SN~1997ef could represent a less extreme case closer in properties to
the known SNe Ic.

At one year, despite its early marked peculiarity, SN~1998bw is
practically indistinguishable from known type Ib (see
Fig.~\ref{fig:final}). Even the high expansion velocities measured
during the first 6 months have slowed down to the values that are
typical for other type Ib SNe ($\sim$5000 km s$^{-1}$). But, the much
higher ejected mass estimated by the models and the high luminosity,
which persists also at these advanced phases (SN~1998bw is 3 mag
brighter than SN 1996N at late phases), tend to support the idea of a
hyper-energetic event.

\begin{figure}
\centering
\includegraphics[height=12cm,angle=-90]{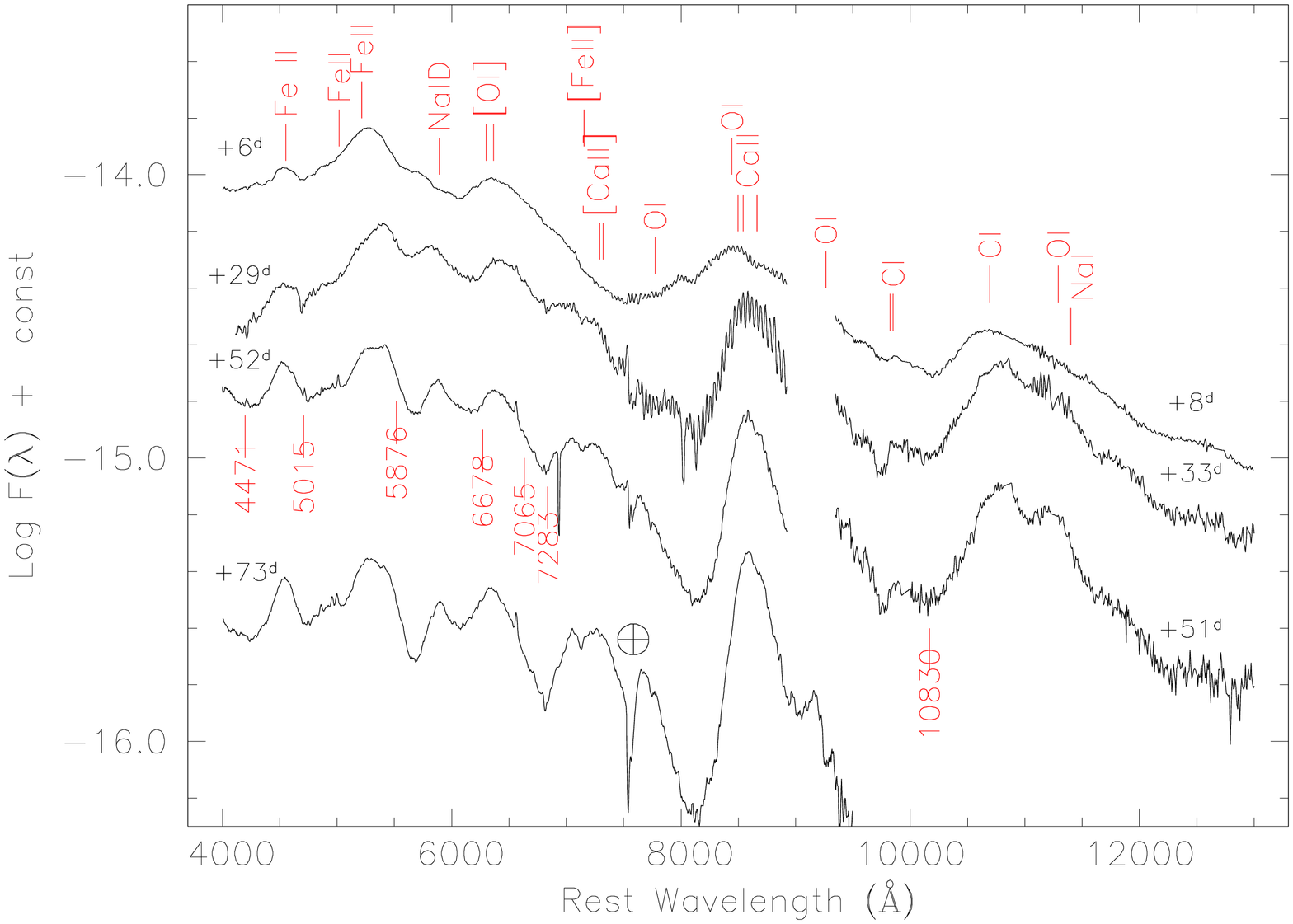}
\caption{ Optical and IR spectra of SN~1998bw   
at comparable phases.  Line identifications from spectral modeling are
plotted for the most prominent emission features (top) and for the
He~I lines (bottom). The He marks are placed at the expected
absorption positions for an expansion velocity of 18,300 km s$^{-1}$.}
\label{fig:helium}
\end{figure}

An important aspect, which may give some hints about the progenitor's
nature is the presence/absence of helium. The optical spectra have
shown no traces of this element (see Fig.~\ref{fig:helium}) and this
is why the SN was classified as a Ic. On the other hand, near-IR
spectroscopy (1.0$-$2.5 $\mu$m) during the early phases has shown 
the presence of a strong emission accompanied by a P-Cyg profile,
which might be He~I 1.08 $\mu$m. This identification is somewhat
supported by the detection of another He~I line at 2.06 $\mu$m.
However, alternative identifications are possible (see the discussion
in Patat et al. \cite{patat01}) and, therefore, the detection of
helium in the spectra of SN~1998bw is not so firm.

\section{A new beginning}

After finishing the work on SN~1998bw I had the impression that the
show was over and we had  met just another peculiar object with no
future. And, more depressing, we were left with more questions than
answers. But {\it nature is subtle} and a very recent GRB, 
named 030329, has shown clear traces of an underlying SN spectrum,
indeed similar to that of SN~1998bw (Garnavich et al. \cite{garnavich};
see also the contribution by T. Matheson, these proceedings).
Even though GRB~030329 was among the brightest ever recorded (while
GRB~980425 was extremely weak), the spectral resemblance to SN~1998bw is 
really astonishing.

\begin{figure}
\centering
\includegraphics[height=8cm]{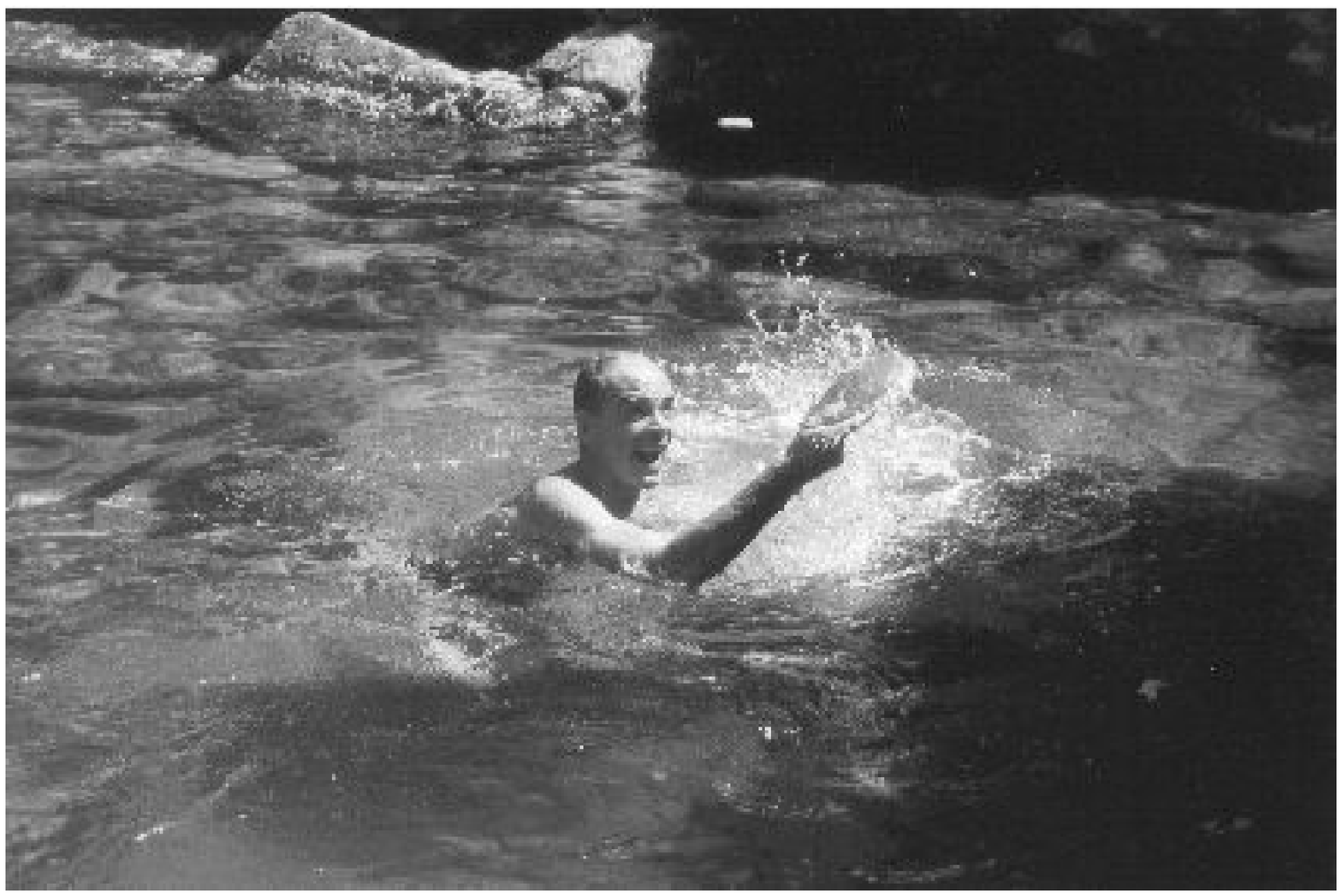}
\caption{Stirling Colgate at Aiguablava (Spain) in 1995. Photo by the author.}
\label{fig:8}
\end{figure}

This clearly indicates that, at least some GRB events are linked to
core-collapse SNe and it confirms the ideas that were born with the
discovery of GRB~980425/SN~1998bw and which have their original seed in the 
pioneering work by Bloom et al. \cite{bloom}.

I have started this review quoting the paper by Clayton, Colgate \& Fishman
\cite{clayton}. Actually, a few years later, in 1974, S. Colgate advanced the
idea that $\gamma$-ray pulses of cosmic origin observed from the Vela
spacecraft could be originated by the core-collapse of massive stars
in distant galaxies (Colgate \cite{colgate}).

Thirty years after, this prophecy seems to come true\footnote{The
first idea actually dates back to 1959, when S. Colgate gave a talk in
Geneva to the Russian delegation of the conference for the cessation
of nuclear testing in space, suggesting that ``SNe or something like
them might trigger our treaty detectors in orbit, causing us to lob
nuclear weapons at each other'' (private communication).}.

%
%
%
%
%

%
%



\printindex
\end{document}